\documentclass[reprint,amsmath,amssymb,aps,pre,abbrv,superscriptaddress,nofootinbib,longbibliography]{revtex4-1}

\usepackage{cancel}
\usepackage{mciteplus,physics,mathtools,comment}
\usepackage{amssymb,graphicx,cancel,amsmath,float,dsfont}
\usepackage[caption=false]{subfig}
\usepackage{dcolumn}
\usepackage{bm}
\usepackage{xcolor}
\usepackage[normalem]{ulem}
\usepackage{braket}
\graphicspath{{Figures/}}
\usepackage[capitalise]{cleveref}
\crefname{appsec}{Appendix}{Appendices}

\newcommand{\iu}{\mathrm{i}\mkern1mu}
\def\e{\mathrm{e}}
\def\ce{\mathrm{ce}}
\def\se{\mathrm{se}}

 \newcommand{\tchi}{\chi}

\newcommand{\tTil}{\tau}

\newcommand{\citer}[1]{Ref. \cite{#1}}

\begin{document}
\title{Balancing long-range interactions and quantum pressure: solitons in the HMF model}
\author{Ryan Plestid}
\email{plestird@mcmaster.ca}
\affiliation{\mbox{Department of Physics and Astronomy, McMaster University, 
1280 Main St. W., Hamilton, Ontario, Canada, L8S 4M1}}
\affiliation{Perimeter Institute for Theoretical Physics, 31 Caroline St. N.,
  Waterloo, Ontario, Canada, N2L 2Y5}
\author{D.H.J. O'Dell}
\email{dodell@mcmaster.ca}
\affiliation{\mbox{Department of Physics and Astronomy, McMaster University, 
1280 Main St. W., Hamilton, Ontario, Canada, L8S 4M1}}
\begin{abstract}
  The Hamiltonian Mean Field (HMF) model describes particles on a ring interacting via a cosine interaction, or equivalently, rotors coupled by infinite-range XY interactions. Conceived as a generic statistical mechanical model for long-range interactions such as gravity (of which the cosine is the first Fourier component), it has recently been used to account for self-organization in experiments on cold atoms with long-range optically mediated interactions. The significance of the HMF model lies in its ability to capture the universal effects of long-range interactions and yet be exactly solvable in the canonical ensemble. In this work we consider the quantum version of the HMF model in 1D and provide a classification of all possible stationary solutions of its generalized Gross-Pitaevskii equation (GGPE) which is both nonlinear and nonlocal. The exact solutions are Mathieu functions that obey a nonlinear relation between the wavefunction and the depth of the meanfield potential, and we identify them as bright solitons.  Using a Galilean transformation these solutions can be boosted to finite velocity and are increasingly localized as the meanfield potential becomes deeper. In contrast to the usual local GPE, the HMF case features a tower of solitons, each with a different number of nodes.  Our results suggest that long-range interactions support solitary waves in a novel manner relative to the short-range case.
  
\end{abstract}

\maketitle


\section{Introduction\label{sec:Intro}} Solitary waves are one of the most distinctive consequences of nonlinearity  and result from a balance between dispersion and nonlinear forces. Their defining property is shape-preserving (i.e.\ dispersionless)  propagation, as first noted  in 1834 by J. Scott Russell when he observed them on the Edinburgh and Glasgow Union Canal \cite{Russell1844}.  The phenomenon has subsequently been extensively studied in classical hydrodynamics \cite{Drazin1989,Tanaka1986,Wang2016,Duan2018}, and also in light propagating in optical fibers \cite{Mollenauer1980,Gordon1983,Kodama1987,Hasegawa1989,Kivshar1998,Mollenauer2006}, and Bose-Einstein condensates (BECs) formed in dilute atomic gases.  In the latter case, quantum pressure (quantum zero point motion) provides the stabilizing dispersion against collapse. Bright  \cite{Strecker2002,Khaykovich2002,Cornish2006,Marchant13,McDonald14,Medley2014}, dark \cite{Burger1999,Denschlag2000,Dutton2001,Jo2007,Engels2007,Weller2008,Stellmer2008,Chang2008} and dark-bright \cite{Becker2008,Hamner2011} varieties of solitary wave have all been observed in BECs.  

In integrable systems solitary waves are guaranteed to survive collisions with one another, and are then referred to as \emph{solitons}, i.e.\ elastically scattering and shape preserving wave packets.  Well-known examples occur in the Korteweg-de Vries equation \cite{Scott1973,Scott2003,Klein2014}, Sine-Gordon equation \cite{Scott1973,Scott2003,Hudak2018} and the nonlinear Schr\"{o}dinger/Gross-Pitaevskii equation (GPE) \cite{Pethick2002,Scott2003,Klein2014}. The nonlinearity in all these wave equations appears as a local term that only depends on the wavefunction at the same point. For example, the GPE has a cubic term  $g \Psi (\mathbf{x},t) \vert \Psi (\mathbf{x},t) \vert^2$ where $g$ parameterizes the sign and magnitude of the self-coupling.  However, there exist physical systems where long-range interactions (LRI) lead to a \emph{nonlocal} nonlinearity  that can also support solitary waves. This is the case in non-neutral plasmas \cite{Davidson2001}, where there is a net Coulomb $1/r$ interaction, in the Calogero-Sutherland (CS) model \cite{Calogero1971,Sutherland1971,Sutherland1972}, where particles interact via a $1/r^2$ potential, and in dipolar BECs \cite{Santos2000,Griesmaier2005,Lahaye07,Koch08,Beaufils08,Lahaye09,Lu11,Aikawa2012}, where dipole-dipole interactions lead to $1/r^3$ interactions. Nonlocal interactions also occur in optical systems, such as those mediated by thermal conduction \cite{Nikolov2003,Krolikowsk04,Nikolov2004,Kong2010,Chen2014}, and their consequences have been observed experimentally \cite{Dreischuh06,Conti09}.  The CS model is integrable and hence supports true solitons \cite{Polychronakos1995,Bardek1995,Sen1997,Bardek2008}, whereas non-neutral plasmas are not integrable systems and only support solitary waves. Dipolar BECs display an instability where the attractive part of the interaction can cause the system to collapse \cite{Koch08}; far from the instability solitary waves are predicted to collide elastically and thus behave as solitons, whereas  close to the instability the collisions become inelastic due to the emission of phonons \cite{Edmonds2016}.  In the BEC literature it is common not to distinguish solitons from solitary waves and thus we use the terms somewhat interchangeably in this paper.

In dipolar BECs solitons are usually analyzed using a generalized Gross-Pitaevskii equation (GGPE) which is an integro-differential equation that incorporates the nonlocal nonlinearity through a Hartree-type mean field term of the form $\Psi(\mathbf{x},t) \int V(\mathbf{x}-\mathbf{x}') \vert  \Psi(\mathbf{x}',t)\vert^2 \ \mathrm{d}\mathbf{x}'$\cite{Pedri2005,Cuevas09,Baizakov2015,Pawlowski2015,Bland2015,Edmonds2016}. This approach has recently been put on a rigorous mathematical footing \cite{Triay18,Eychenne18}; for reviews of nonlocal nonlinear Schr\"{o}dinger equations we refer the reader to References  \cite{Ablowitz1} and \cite{Ablowitz2}. It has also been suggested that the Manakov equations \cite{Manakov1974} (which describe  both two component BECs \cite{Kevrekidis2016} and randomly birefringent light in optical fibers \cite{Evangelides1992,Wai1994}) can give rise to a so-called algebraic nonlinearity provided the system is prepared with a specific set of initial conditions \cite{Yang2018}. Integral terms can also appear in the equations describing the motion of vortices in superfluids at finite temperatures; they arise, for example, from the mutual friction between a solitary wave along a vortex and its surrounding flow \cite{vanGorder2015}.

Another physically important class of nonlocal nonlinearity is found in self-gravitating systems. In particular, compact yet stable astrophysical objects made of bosons and known as ``Bose stars'' have been hypothesized \cite{Jetzer1992,Schunck2003}. These can be identified as solitons if the attraction due to gravity is balanced by quantum pressure \cite{Seidel94}. The realization that dark matter in the universe may be bosonic and cold enough to Bose condense into such Bose stars has driven considerable interest in these systems \cite{Khlebnikov00,Elgart2007,Sikivie09,Sikivie12,Schive14,Levkov2017,Levkov2018}. For the most part these studies use the non-relativistic Schr\"{o}dinger-Newton (also known as the Schr\"{o}dinger-Poisson) equations \cite{Moroz1998,Bahrami2014} which result in a GGPE similar in form to both the GGPE used for dipolar BECs and also the equation to be studied in this paper. We also note in passing that laboratory analogues of the Schr\"{o}dinger-Newton system have been proposed in the form of atomic BECs with a $1/r$ interaction provided by laser-induced dipole-dipole interactions \cite{Odell00,Giovanazzi01}.

The focus of the present paper is the Hamiltonian Mean Field (HMF) model \cite{Antoni1995} which describes $N$ particles of mass $m$  living on a ring. They have positions $\theta_i\in (-\pi,\pi]$, angular momenta $L_{i}$, and interact via a pairwise cosine potential of strength $\epsilon$ giving the Hamiltonian
\begin{equation}
  H=\sum_i \frac{L_{i}^2}{2I}+ \frac{\epsilon}{N} \sum_{i<j} \cos(\theta_i-\theta_j),
  \label{eq:Hclass}
\end{equation}
where $I=mR^2$ is the moment of inertia for a ring of radius $R$. Since every particle interacts with every other due to the long-range nature of the interactions, a factor of $1/N$ is explicitly included to make the energy extensive (often  termed the  Kac prescritpion \cite{Kac1963}). The cosine potential can be thought of as the first non-constant term in a Fourier series expansion of a gravity- or Coulomb-like $1/r$ interaction around the ring, but without the singularity at $r=0$ that otherwise complicates the treatment of such potentials.  Another way to view the HMF model is as a system of $N$ rotors interacting via an infinite range XY interaction.  If $\epsilon >0$ the particle interactions are repulsive at small distances, or equivalently, the rotors experience an antiferromagnetic coupling and hence prefer to anti-align. If instead $\epsilon < 0$ the interactions are attractive/ferromagnetic and the rotors prefer to align. The average magnetization of the rotors along an axis specified by its angle $\varphi$ to the vertical is given by $\langle \cos (\theta-\varphi) \rangle = (1/N) \sum_{i} \cos(\theta_{i} -\varphi)$, and this quantity serves as an order parameter for a symmetry breaking phase transition (clustering transition) which occurs in the attractive case at low temperatures \cite{Dauxois2002}. In this paper we focus on the attractive/ferromagnetic case.

The HMF model was originally written down as a toy model and its significance lies in the fact that it is simple and yet able to capture many of the general qualitative properties of systems with LRI, for reviews see \cite{Dauxois2002,Campa2009,Levin2014}.  However, more recently it was realized that cold atomic gases  trapped in high-finesse optical cavities can directly realize the HMF model [up to an additional term of the form $\cos (\theta_{i}+\theta_{j})$]  \cite{Schutz2014}. Here, the long-range interactions are electromagnetic in origin and mediated  by optical cavity modes that can extend over the entire gas. On going experiments \cite{SelfOrgVuletic03,DickePT2010,Roton2012,Supersolid2017,ControlOrderHemmerich2018,SpinorSelfOrdering_Lev_Lab_2018}, many with BECs, have demonstrated  symmetry breaking and self-organization in the atomic density distribution that is described by the HMF model  \cite{Schutz2014,Schutz2015,Schutz2016,Keller2017,Keller2018}. 

One of the most striking properties of systems with LRI has long been appreciated by the astrophysics community: the two-body relaxation time (also known as the Chandrasekhar relaxation time) to thermodynamic (Boltzmann-Gibbs) equilibrium,  diverges with the number of particles $t_{\mathrm{relax}} \sim N t_{\mathrm{cross}}/10 \log[N]$, where $t_{\mathrm{cross}}$ is the typical time for a particle to cross the system \cite{Binney2011}. Thus, in the thermodynamic limit $N \rightarrow \infty$ the system never achieves thermodynamic equilibrium.  Nevertheless, when a self-gravitating system is disturbed from equilibrium the common mean field potential that arises from the long-range nature of gravity becomes time dependent and can drive a rapid, collisionless relaxation mechanism known as violent relaxation whose timescale does not depend on the number of particles. This efficiently mixes phase space \cite{Binney2011}, but the process is non-ergodic and the resulting quasi-stationary state is not the equilibrium state predicted by the microcanonical ensemble. However, coarse graining of the phase space distribution function by a macroscopic observer averages over the increasingly fine structures that develop during collisionless relaxation and in this way conventional statistical mechanics approaches can be applied  \cite{Lynden-Bell1967,Yamaguchi2008,Pakter2011,Levin2014}. The HMF model has been shown to display violent relaxation \cite{Dauxois2000,Leyvraz2002,Barre2002,Barre2002a}, as well as other generic consequences of LRI, including spontaneous symmetry breaking in low dimensions, i.e.\  the clustering phase transition mentioned above   (LRI violate the Mermin-Wagner theorem), and the so called ``core-halo'' statistics observed at late times in gravitational dynamics simulations \cite{Yamaguchi2008,Pakter2011,Levin2014}.

The HMF model can be extended to describe quantum systems with LRI if we replace the kinetic energy term in \cref{eq:Hclass} by its quantum operator $-\frac{\hbar^2}{2mR^2}\partial_{\theta_i}^2$, and the system is then equivalent to an infinite range O(2) quantum rotor model \cite{Ye1993,KTK1994,Read1995,Kennett2001,Sachdev2011}. Motivation for studying the quantum problem comes both from cold atom experiments and the Bose star picture of dark matter mentioned above.  The equilibrium states of the quantum HMF model have been examined in \cite{Chavanis2011a,Chavanis2011} and the dynamics, including violent relaxation, were recently studied in \cite{Plestid2018} where it was found that the automatic coarse-graining of phase space at the level of Planck's constant $h$ can strongly modify the relaxation in the deep quantum regime.  Furthermore, in its quantum form the HMF model bears a resemblance to the CS model mentioned above, which, when defined on a finite domain with periodic boundary conditions, has a pairwise interaction $V(\theta_i-\theta_j)=1/\sin^2(\theta_i-\theta_j)$. Like the HMF model, the CS model has a periodic infinite range interaction. This connection is relevant in the present context because the CS model supports true solitons. The HMF model is not thought to be exactly integrable, but intriguingly, classical long-range interacting many-body systems are known to be described (exactly in the $N\rightarrow \infty$ limit \cite{Hepp1977}) by the Vlasov equation. Vlasov dynamics \emph{are integrable} for a one dimensional system \cite{Ogawa2012}, such as the HMF model\footnote{Furthermore, the infinite set of Casimir invariants for the Vlasov equation can mimic the effects of integrability for higher dimensional systems with LRI see, e.g., \cite{Patelli2014}.}. Furthermore, the classical (i.e.\ $\hbar\rightarrow0$) limit of the GGPE corresponds to the ``zero-temperature'' limit of the Vlasov equation \cite{Barre2002,Barre2002a,Plestid2018}. Therefore, although the HMF model is not exactly integrable, its classical dynamics are nearly integrable due to the structure of the Vlasov equation. We speculate that the ``pseudo-integrability'' extends also to the GGPE and may allow the solitary waves (bright solitons) presented here to scatter elastically.

Our goal in this work is to study solitary waves in the quantum HMF model with attractive/ferromagnetic interactions. We focus on the model's associated GGPE, appropriate for describing the dynamics of a Bose condensed state (the HMF model, despite its name, describes a many-body system: to obtain a GGPE we assume all the particles occupy the same quantum state). We find that the exact solutions to this equation are Mathieu functions that satisfy a nonlinear self-consistency relation.   As the interaction strength tends to infinity the number of these solutions also becomes infinite and we identify them as bright solitons each with a different number of nodes. Both the large number of solutions and the fact they have nodes makes them  unusual when compared to the standard local GPE case \cite{Pethick2002}. We attribute these differences to the LRI themselves and hypothesize that this might be a generic feature. 

This paper is organized as follows: In \cref{sec:Hartree} we introduce the GGPE for the HMF model. In \cref{sec:selfconsistent} we show how to find the full set of exact stationary solutions to the GGPE via a self-consistent Mathieu equation, and in \cref{sec:boost} we boost these solutions to finite velocity to obtain traveling waves. In \cref{sec:Solitons} we explore the regime in which the solutions can be considered as bright solitons, and discuss the parametric dependence of the stationary solutions on Planck's constant. Next we discuss the asymptotic behavior of these solutions and derive useful analytic expressions. In \cref{sec:Stability} we study their stability at leading order in the strong coupling regime by linearizing the equations of motion and analyzing the mode spectrum. Finally, in \cref{sec:Conclusions} we summarize our work and discuss future directions of investigation.


\section{Generalized Gross-Pitaevskii Equation \label{sec:Hartree}} 
An atomic BEC with short-range interactions is described by the standard GPE with a local cubic nonlinearity \cite{Pitaevskii2003}
\begin{equation}
  \iu \hbar \frac{\partial \Psi}{\partial t} =\left[-\frac{\hbar^2}{2m}\nabla^2 +V_{\mathrm{ext}}(\mathbf{x})+ g N \vert \Psi \vert^2  \right] \Psi
  \label{eq:GPEStandard}
\end{equation}
where $N$ is the number of atoms, $\Psi(\mathbf{x},t)$ is the condensate wavefunction normalized to unity: $\int_{-\infty}^{\infty} \vert \Psi \vert^2 \mathrm{d} \mathbf{x} =1$, $V_{\mathrm{ext}}(\mathbf{x})$ is a possible external potential, e.g.\ a harmonic trap or periodic optical lattice potential, and the coupling constant $g$ parameterizes the interatomic interactions  (usually of the van der Waals-type). Any stationary solution to this equation can be found in the usual way by putting $\Psi(\mathbf{x},t)=\psi(\mathbf{x}) \exp [- \mathrm{i} \mu t/\hbar]$, giving 
\begin{equation}
\mu \psi =\left[-\frac{\hbar^2}{2m}\nabla^2 +V_{\mathrm{ext}}(\mathbf{x})+ g N \vert \psi \vert^2  \right] \psi \ .\label{eq:GPEstatic}
\end{equation}
If this were a linear Schr\"{o}dinger equation the eigenvalue $\mu$ would be the energy $E$ of the state $\psi$. However, due to the nonlinearity of the GPE $\mu$ is in fact the chemical potential which is the change in energy associated with adding a particle to the system: $\mu=\partial E /\partial N$ \cite{Pitaevskii2003}. In a waveguide configuration, where the BEC is tightly trapped in the $x$ and $y$ directions but untrapped along the $z$ direction, the problem becomes effectively one-dimensional. Specializing further to the case of attractive interactions ($g<0$), the stationary GPE  has the bright soliton solution
\begin{equation}
\psi_{0}(z)=\psi_{0}(0) \frac{1}{\cosh(z/\sqrt{2} \xi)}
\label{eq:fundamental}
\end{equation}
where $\rho_{0}=\vert \psi_{0}(0) \vert^2$ is the central density and $\xi=1/\sqrt{2 m \vert g \vert \rho_{0}}$ is a characteristic length called the healing length \cite{Pitaevskii2003}. In the optical soliton literature this solution is called the \textit{fundamental soliton} and the coordinate $z$ is replaced by $z-vt$ representing a shape-preserving waveform propagating at the group velocity $v$ (there is also a multiplicative phase factor we shall not specify here) \cite{Mollenauer2006}.

 By contrast, the LRI in the HMF model lead to a \emph{nonlocal nonlinearity} 
 \cite{Chavanis2011,Plestid2018}. The HMF model lives on a ring so that the wavefunction obeys periodic boundary conditions $\Psi(\theta,t)=\Psi(\theta+2\pi,t)$ and it also does not include an explicit potential $V_{\mathrm{ext}}(\mathbf{x})$.  Fixing the interactions to be attractive (i.e.\ $\epsilon<0$),  the HMF model's GGPE can be written in reduced variables as  
\begin{subequations}\label{eq:GGPE}
  \begin{equation}
    \iu \tchi \pdv{\Psi}{\tTil}=\qty[-\frac{\tchi^2}{2}\pdv[2]{\theta} - \Phi(\theta,\tau)]\Psi
    \label{eq:GGPE-a}
  \end{equation}
  \begin{equation}
    \mbox{where   } \quad
    \Phi(\theta,\tau)=\int_{-\pi}^{\pi}\rho(\theta',\tau)
    \cos(\theta-\theta') \ \mathrm{d}\theta' 
    \label{eq:GGPE-b}
  \end{equation}
\end{subequations}
is a nonlocal Hartree potential that depends on the integral  over the angular probability density $\rho(\theta,\tau)= \vert \Psi(\theta, \tau) \vert^2$.
We have introduced the  quantities $\tau=t\times\sqrt{\frac{|\epsilon|}{mR^2}}$, and $\tchi$, the latter of which plays the role of a dimensionless Planck's constant:
\begin{equation}
\tchi=\frac{\hbar}{\sqrt{m R^2 \abs{\epsilon}}}  \,  .
\label{eq:tchi_defn}
\end{equation}
Note that $\tchi$ depends on the magnitude of the interaction strength $\abs{\epsilon}$. 

The seemingly complicated integro-differential equation given in \cref{eq:GGPE} can be simplified by expanding the probability density in its Fourier components $\rho(\theta')=\sum \hat{\rho}_k \e^{\iu k \theta'}/2\pi $. We see that $\Phi$ depends only on $\hat{\rho}_{\pm 1}$, which we can write in all generality as $\hat{\rho}_{\pm 1}:=M(\tau) \exp\qty[\mp \iu \varphi(\tau)]$. From this it follows that the Hartree potential takes the remarkably simple form
\begin{equation}
  \Phi\qty(\theta,\tTil)=M\qty(\tTil)\cos[\theta-\varphi\qty(\tTil)]  \ .
  \label{eq4:MFSE}
\end{equation}

The physical significance of this result is that all the individual two-body XY potentials of the many-body theory are replaced in the Gross-Pitaevskii theory by a single collective potential $\Phi\qty(\theta,\tTil)$ which retains the same XY form (a cosine) but breaks the angular symmetry by picking out a particular direction specified by $\varphi(\tau)$ along which the magnetization is $M(\tau)$.  Furthermore, because $M(\tau)$ is the coefficient of the $\pm 1$ terms in the Fourier expansion of the probability density, we can always project it out from the angular probability distribution via the integral
\begin{equation}
M(\tau) =\int_{-\pi}^{\pi} \rho(\theta,\tau) \cos(\theta - \varphi) \ \dd \theta
\label{eq:magnetization}
\end{equation}
which is the continuous version of the definition of the magnetization  given in the Introduction,  $M = \langle \cos (\theta-\varphi) \rangle = (1/N) \sum_{i} \cos(\theta_{i} -\varphi)$.

The stationary states $\Psi(\theta,\tau)=\psi(\theta)\exp[-\iu \mu\tau/\tchi]$ of the GGPE satisfy the equation
\begin{equation}
  -\frac{\tchi^2}{2}\pdv[2]{\psi}{\theta} + \qty(-  \mu -M[\psi] \, \cos \theta)\psi=0
  \label{eq:NL-Mathieu}
\end{equation}
where we have taken advantage of the fact that in the stationary case we can define our coordinates such that $\varphi=0$. As for the case of contact interactions, the eigenvalue $\mu$ is not the same  as the energy $E=\langle \psi \vert H \vert \psi \rangle $ of the state $\psi$ associated with Hamiltonian given in \cref{eq:Hclass}, but is the chemical potential.
 
The stationary GGPE given in \cref{eq:NL-Mathieu} has the same form as the Mathieu equation \cite{Olver2010} 
 \begin{equation}
   \pdv[2]{w}{z} + \qty[a-2q\cos(2z)]w=0~,
   \label{eq:Mathieu}
 \end{equation}
whose solutions are Mathieu functions. These are denoted $\textrm{ce}_n(z;q)$ and $\textrm{se}_n(z;q)$ and have eigenvalues $a=A_n(q)$ and $B_n(q)$, respectively. In general, Mathieu functions also depend on a second parameter, which physically is the quasimomentum. However, here the quasimomentum is fixed to be zero by the periodic boundary conditions imposed by the ring. 

There is a crucial difference between the standard Mathieu equation and the GGPE.  Whereas the former corresponds to a linear problem where the ``depth parameter'' $q$ of the cosine potential takes a fixed specified value, in our problem the depth of the cosine potential is the magnetization $M[\psi]$ which is a functional of the wavefunction and so depends on the solution itself. In other words, we have a nonlinear eigenvalue problem which, remarkably, has eigenvectors that are Mathieu functions obeying a linear equation but which must be supplemented by the self-consistency condition
\begin{equation}
  M[\psi]=\int_{-\pi}^{\pi} \big|\psi(\theta')\big|^2 \cos(\theta') \ \dd \theta' \ .
  \label{eq:mag-self-cons}
\end{equation}
We note that this is simply a restatement of \cref{eq:magnetization} for a density $\rho(\theta)=|\psi(\theta)|^2$ that is independent of time.

An analogous situation occurs in cavity-QED where atoms are trapped in an optical cavity pumped by a laser \cite{Prasanna2009,Prasanna2011,Prasanna2013,kessler2016,Georges2017,Lee2017}. The laser light forms a standing (or travelling \cite{Goldwin2014,Samoylova2015}) wave inside the cavity which the atoms experience as a sinusoidal potential via the optical dipole interaction. The atoms' centre-of-mass wavefunction is therefore also determined by the Mathieu equation. However, the interaction acts back on the light which sees the atoms as a refractive medium. This backaction shifts the cavity's resonance frequency, and hence controls the amount of laser light that can enter the cavity, by an amount that depends on the overlap between the atomic density distribution and the optical mode. In this way the atomic density profile affects the depth of the sinusoidal potential which in turn affects the atomic density profile. The problem is therefore nonlinear and also leads to a Mathieu equation with a parameter $q$ that must be determined self-consistently from the atomic wavefunction like in \cref{eq:mag-self-cons}. One effect of this nonlinearity is the appearance of curious loops in the band structure that are not present in the linear problem and which can lie in the band gaps \cite{Prasanna2011,Coles2012}. 

Band gap loops also occur in the problem of a BEC in an optical lattice of fixed depth, i.e.\ \cref{eq:GPEStandard} with $V_{\mathrm{ext}}(x)=V_{0} \cos (kx)$, where the nonlinearity comes purely from interatomic interactions modelled by the cubic nonlinearity. This situation has been investigated both experimentally \cite{Eiermann04} and theoretically \cite{Wu00,Bronski01,Wu03,Machholm03,Louis03,Machholm04} where it is found that the loops correspond to two different types of solutions:  periodic trains of solitons, i.e.\ spatially extended solutions  \cite{Machholm03,Machholm04}, and localized band gap solitons \cite{Louis03}. Despite their localization, these latter solitons can have one or more nodes.  In the HMF problem we also have a cosine potential but it is limited to a single period by the periodicity of the ring. This fixes the quasimomentum to zero and hence collapses the bandstructure to the centre of the Brillouin zone.  Still, we shall find analogous solutions to localized band gap solitons as will be described below.


\section{Self-consistent Mathieu functions \label{sec:selfconsistent}}

To convert between the GGPE given in \cref{eq:NL-Mathieu} and the standard form of the Mathieu equation  given in \cref{eq:Mathieu} we make the identifications\footnote{The shift in coordinates is equivalent to a negative value of q and accounts for the attractive nature of the interparticle interaction.} $\theta=2(z+\pi/2)$ and  $\partial_z^2=  4\partial_\theta^2$.  Multiplying both sides of \cref{eq:Mathieu} by a factor of $-\tchi^2/8$ we find 
\begin{equation}
\mu=\frac{\tchi^2}{8}a,\quad  \text{and} \quad  M=\frac{\tchi^2}{4}q \label{eq:Self-Consistent}.
\end{equation} 
The solutions we require are the ones that are $2 \pi$-periodic in the angle $\theta$, and these correspond to $\textrm{ce}_{2n}$ and $\textrm{se}_{2n}$ with $n\geq0$ an integer.

\begin{figure}[t]
\includegraphics[width=\linewidth]{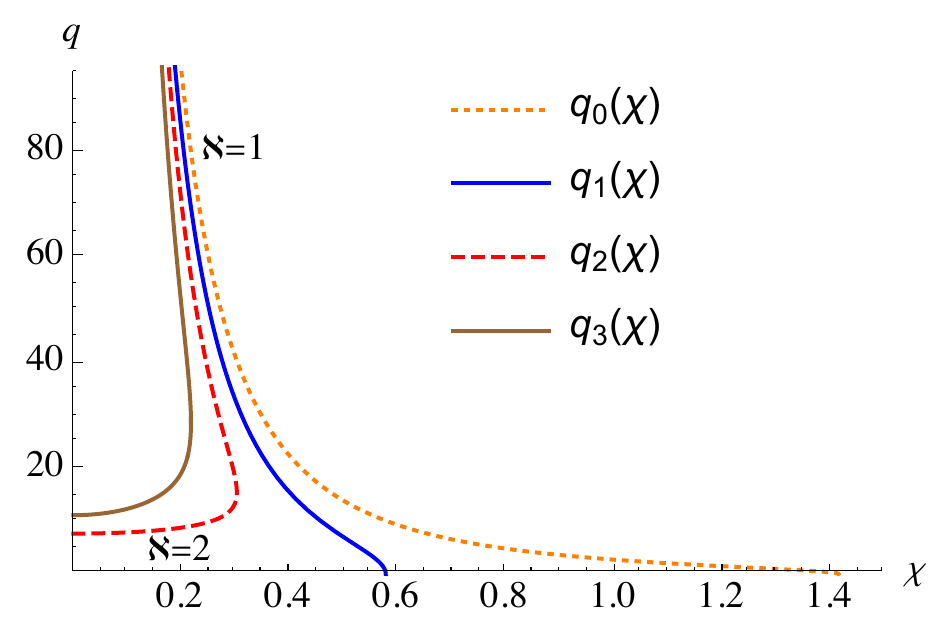}
\caption{Self consistent values of the potential depth, as parameterized by $q$, as a function of $\tchi$ for the first four stationary solutions: this plot shows which Mathieu functions satisfy both \cref{eq:NL-Mathieu} and the self consistency condition \cref{eq:mag-self-cons}. Using \cref{eq:psi-n-def} we can then find explicit expressions for the stationary state $\psi_n(\theta;\tchi,\aleph)$. The parameter $\aleph$ labels the different branches of $q_n(\tchi;\aleph)$. Note that both $n=2$ and $n=3$ have two branches as the function turns back on itself. This is a generic feature for $n\geq 2$.  }
\label{fig:q-of-chi}
\end{figure}

To find self-consistent solutions of the GGPE we use the following algorithm:
\begin{enumerate}
\item We first treat $q$ as a fixed parameter like in the usual linear theory of Mathieu functions. Taking a given Mathieu function of fixed $n$ and $q$ (which we denote $q_n$) we compute $M(q_n)$ using equation \cref{eq:mag-self-cons} where $\psi=\text{ce}_n(z;q_n)/\sqrt{\pi}$ or $\psi=\text{se}_{n+1}(z;q_n)/\sqrt{\pi}$ and $z=(\theta-\pi)/2$. 
\item Next, we obtain $\chi(q_n)$ by using \cref{eq:Self-Consistent} which gives $\chi(q_n)=\sqrt{4 M(q_n)/q_n}$. 
\item The above two steps are repeated a large number of times for different values of $q$ to obtain a map between $\chi$ and $q$. This must be done separately for each Mathieu function (each value of $n$).
\item Although we have treated $q$ as a parameter upon which $\chi$ depends, in reality the situation is reversed with $q_n$ being determined self-consistently in terms of $\chi$. We therefore invert the map $\chi(q_n)$ to find $q_n(\chi ; \aleph)$ where we have introduced the integer $\aleph$ to label different branches of the function in the case that $\chi(q_n)$ is not invertible. The results are shown in Figure \ref{fig:q-of-chi}.
\end{enumerate}

\begin{figure}[t]
  \includegraphics[width=0.5\linewidth]{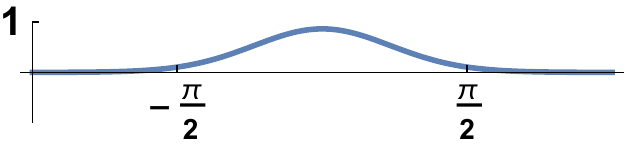}~
  \includegraphics[width=0.5\linewidth]{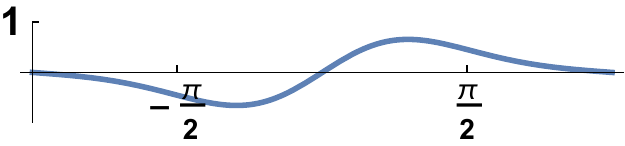}\\
  \caption{Plots of the two self-consistent stationary solutions to the GGPE [Eqns.\ (\ref{eq:GGPE-a}) and (\ref{eq:GGPE-b})] that exist at $\tchi=0.5$. They are the $n=0$ and $n=1$ states, where $n$ gives the number of modes. The solutions are periodic with period $2 \pi$.}
  \label{fig:tinystationarystates}
\end{figure}

\begin{figure}[h]
  \includegraphics[width=0.5\linewidth]{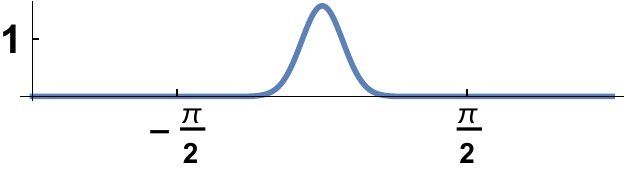}~
  \includegraphics[width=0.5\linewidth]{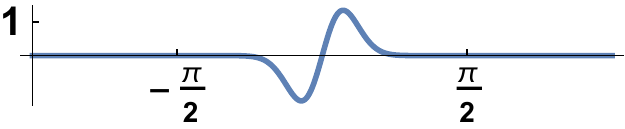}\\
  \includegraphics[width=0.5\linewidth]{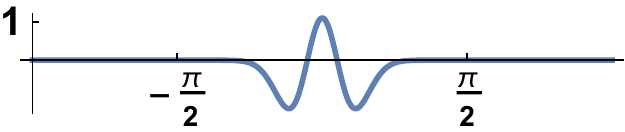}~
  \includegraphics[width=0.5\linewidth]{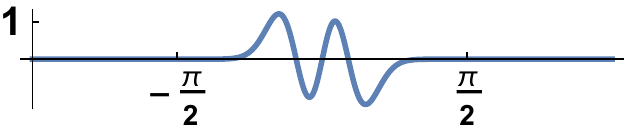}\\
  \includegraphics[width=0.5\linewidth]{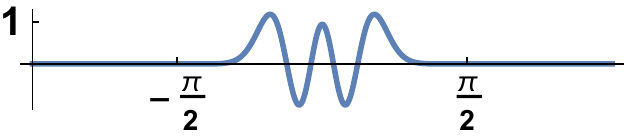}~
  \includegraphics[width=0.5\linewidth]{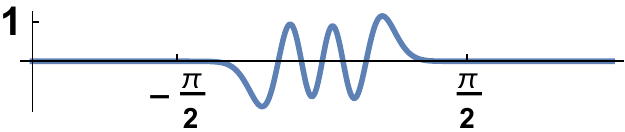}\\
  \caption{Plots of the first six self-consistent stationary solutions to the GGPE for $\tchi=0.05$. These solutions all lie on the 
  upper branch of \cref{fig:q-of-chi},  i.e.\ $\psi_n(\theta ; \tchi, \aleph=1)$ where $n\in\{0,1,2,3,4,5\}$. Note that even for moderately small values of $\tchi$ the first few stationary states are quite well localized, and upon a Galilean transformation to finite velocity can be interpreted as solitary wave solutions. All states are periodic with period $2 \pi$.}
  \label{fig:stationarystates}
\end{figure}

Noting that Mathieu functions are conventionally normalized on the unit circle such that $\int \abs{\textrm{ce}_n(\theta)}^2\dd \theta = \pi $, and $\int \abs{\textrm{se}_n(\theta)}^2\dd \theta = \pi $, we define our stationary states as 
\begin{equation}
  \psi_n(\theta; \tchi,\aleph)=\frac{1}{\sqrt{\pi}}
  \begin{cases}
    \hspace*{0.9em}\textrm{ce}_{n}\qty[\frac{\theta-\pi}{2};q_n(\tchi;\aleph)]\hspace*{-0.2em} & n~\text{even} \\
    \textrm{se}_{n+1}\qty[\frac{\theta-\pi}{2};q_n(\tchi;\aleph)]\hspace*{-0.2em} & n~\text{odd} 
  \end{cases}
  \label{eq:psi-n-def}
\end{equation}
where $q_n(\tchi;\aleph)$ is obtained via the algorithm outlined above. 
In Figs. \ref{fig:tinystationarystates} and \ref{fig:stationarystates} we plot  some examples of these stationary solutions for  $\chi=0.5$ and $\chi=0.05$, respectively, where in the latter case both branches of solutions exist, so we have chosen the  $\aleph=1$ branch.  We see that $n$ gives the number of nodes.

According to its definition in \cref{eq:tchi_defn}, $\tchi$ decreases as the magnitude of the (attractive) interaction strength $\vert \epsilon \vert$ increases, and hence the $\aleph=1$ branch solutions correspond most naturally to bright solitons: from \cref{fig:q-of-chi} we see that as $\tchi$ decreases the potential, as parameterized by $q$, becomes deeper and the states are more tightly bound as can be seen by comparing \cref{fig:tinystationarystates} with \cref{fig:stationarystates} (see \cref{sec:Solitons} for further justification that these are really localized solutions). By contrast, the $\aleph=2$ branch corresponds to shallower potentials which vary only slightly with $\tchi$. This branch continuously passes to negative $\tchi$ (not shown in \cref{fig:stationarystates}) corresponding to repulsive interactions which shows that the repulsive HMF model can also support stationary states, however we emphasize that these states generally have smaller values of $q$ and are less localized than their attractive interacting counterparts. 

The first two solutions, $\psi_0$ and $\psi_1$, only have a single branch and are shown in Fig.\ \ref{fig:tinystationarystates}. For weak interactions ($\tchi \gg 1$) ones finds that $q$ is zero so that the self-consistent Hartree cosine potential is zero. These solutions then correspond to ordinary sine waves (although, as shown in \cref{fig:mu-of-chi}, for $\tchi > \sqrt{2}$ the energy eigenvalue/chemical potential corresponding to $\psi_{0}$ takes the value $\mu=0$ and thus this is a trivial solution corresponding to an infinite wavelength, i.e.\ a flat density profile). However, at a critical value of $\tchi$, which is different for each solution, the Hartree potential switches on and the solutions evolve continuously into Mathieu functions. For $\psi_0$ this occurs at $\tchi=\sqrt{2}$ \cite{Chavanis2011}, and for $\psi_1$ it occurs at $\tchi=1/\sqrt{3}$. The higher solutions behave differently, as can be seen from \cref{fig:q-of-chi}. In their case each solution again switches on below a critical value $\tchi$, but $q$ takes on a finite value at the point each solution appears (and which immediately splits into two branches).

With the knowledge of the dependence of the $q_{n}$'s upon $\chi$ depicted in Fig.\ \ref{fig:q-of-chi} we can straightforwardly obtain physical quantities such as the chemical potentials and magnetizations of all the stationary states. These are plotted in \cref{fig:mu-of-chi} 
and \cref{fig:mag-of-chi}, respectively. Each chemical potential $\mu_{n}=\partial E_{n} /\partial N$ gives the change in energy of its respective state $E_{n}=\langle \psi_{n} \vert H \vert \psi_{n} \rangle$ when a particle is added to the system. From \cref{fig:mu-of-chi} we see that the gaps between the different $\mu_{n}$ vanish linearly as $\tchi \rightarrow 0$, so that all $\mu_{n}$ tend to the common value of -1. However, the rate at which they approach this limit is higher the higher the state so that the largest gap is between $\mu_{0}$ and $\mu_{1}$.

Although the chemical potential is sensitive to the clustering or ordering phase transition that occurs at the critical value of $\tchi$, it is the magnetization which is usually taken as the order parameter for the transition \cite{Dauxois2002}. In particular, examining the behaviour of the magnetization $M_{0}$ of the lowest state in \cref{fig:mu-of-chi}, we see that when the coupling is weak ($\tchi \gg 1$) the magnetization is zero but at the critical value $\tchi=\sqrt{2}$ it begins to take on finite values indicating a second order (continuous) phase transition. Similar behaviour is found for $M_{1}$.  However, the behavior of the magnetization of the higher states is more complicated; nevertheless the $\aleph=1$ branches of these states, which correspond to bright solitons, tend to full magnetization $M=1$ in the strong coupling regime $\tchi \ll 1$.

The results given in this section generalize those obtained by Chavanis \cite{Chavanis2011} for the ground state of the quantum HMF model.
In particular, the approach described above allows one to classify \emph{all possible} stationary states of the GGPE. They are labelled by their number of nodes $n$, and  their depth parameter $q_n(\chi;\aleph)$.  Furthermore, whilst the approach followed in \citer{Chavanis2011} gives an expansion for the small $q$ properties of the ground state $\psi_0(\theta;\tchi)$, as well as the leading order result in the strong-coupling regime $\chi \ll 1$,
the approach followed here is valid for all values of $\tchi$ and, upon computation of $q_n(\tchi)$, provides an analytic expression for $\psi_n(\theta)$.
The standard large and small $q$ asymptotics of the Mathieu functions \cite{Olver2010} can then be used to obtain analytic approximations for $q_n(\tchi;\aleph=1)$.

The fact that the GGPE admits a tower of stationary solutions $\psi_{n}(\theta;\chi,\aleph)$, each labelled by its number of nodes $n$ and branch $\aleph$, is quite distinct from the case of the local GPE where there is only a single stationary bright soliton solution, i.e.\ the nodeless fundamental soliton given in Eq.\ (\ref{eq:fundamental}).  In particular, the GGPE's tower of stationary solutions should not be confused with the so-called ``higher order solitons''  found in the local GPE which can display multiple peaks and nodes at certain instants of time \cite{Satsuma74,Mollenauer1980,Mollenauer2006}.   These higher order solitons are time dependent combinations of the fundamental and do not correspond to the individual stationary solutions we describe above. Indeed, as famously shown by Zakharov and Shabat \cite{Zakharov72}, the inverse scattering transform method can be applied to the local GPE and in general gives rise to a nonlinear superposition of multiple fundamental solitons and continuous waves (see p21 in \cite{Hasegawa1989} and p245 in \cite{Mollenauer2006} for a summary of these results).

\section{Boosted solutions}
\label{sec:boost}

The same solutions as already described for the stationary case can be transformed to travelling waves with velocity $v$
by performing a Galilean boost 
\begin{equation}
\psi_n(\theta)\e^{-\iu \mu_n \tau/\tchi}\rightarrow\psi_n(\theta-v\tau)\e^{\iu v\theta/\tchi }~\e^{-\iu (\mu_n + v^2/2)\tau/\tchi}~.
\label{eq:Boost}
\end{equation}
In order that these wavefunctions still satisfy the periodic boundary conditions we require that
$v/\chi=n$ with $n\in \mathbb{Z}$. The existence and classification of these wavefunctions is the main result of our work. In the next section we will see when they can be considered to be solitary waves.

\section{Emergence of solitons at strong coupling \label{sec:Solitons}}
Having established the existence of non-trivial stationary and traveling waves we will now argue that the crucial solitonic property of localization emerges in the strong-coupling regime $\tchi \ll 1$. Finally, we show how in this same limit an explicit asymptotic series for the depth parameter $q_n(\chi)$ can be obtained.

\subsection{Localization in the strong-coupling regime}
In the limit of strong coupling the Hartree meanfield potential is deep relative to the kinetic energy, and the magnetization can be expected to saturate to unity. It then follows via the relation $q= 4M/\tchi^2$ that this limit corresponds to large values of $q$ and we will see that this is indeed the case. 

As $q\rightarrow \infty$ the eigenvalues, $A$ and $B$, of the Mathieu equation display the well known asymptotic behavior as a function of $q$ \cite{Olver2010}
\begin{equation}
\begin{rcases}
  &A_n(q)\\
  &B_{n+1}(q)
\end{rcases}\sim -2q\qty[1 -\frac{1}{\sqrt{q}}(2n+1) + \order{\frac{1}{q}}]
\end{equation}
from which we can identify that the low-lying states are bound within a deep well\footnote{The spectrum is that of a harmonic oscillator with anharmonic corrections occurring at $\order{1/q}$.} whose minimum is spontaneously chosen around the ring. There are two classical turning points
\begin{equation}
\theta^\pm_\text{turn}\sim \pm\frac{(4n+2)^{1/2}}{q^{1/4}}+\order{\frac{1}{q^{1/2}}}. 
\label{eq:turn-points}
\end{equation}
At distances $\abs{\theta} \gtrsim 2 \theta^+_\text{turn}$ the wavefunction is exponentially suppressed and consequently the state is localized in a region around the meanfield potential's minimum [as can be seen in \cref{fig:stationarystates}]. The size of this region shrinks with $q$ and shows that localized stationary states emerge  in the strong-coupling regime, and, by Galilean invariance, so too do finite-velocity travelling solitons. Thus, we identify the solutions as solitary waves because they can be arbitrarily localized and their shape is determined by a competition between quantum dispersion and a classical meanfield potential, $\Phi$, which is the source of the non-linearity in \cref{eq:GGPE}.

\begin{figure}[t]
  \includegraphics[width=\linewidth]{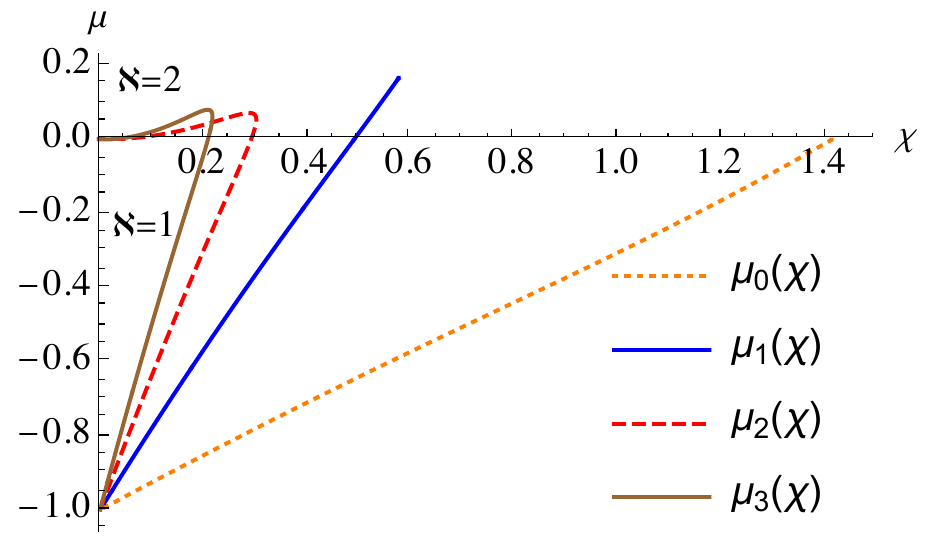}
  \caption{The eigenvalues $\mu$ of the GGPE as a function of $\tchi$ for the first four stationary solutions.  In this figure $\mu_{0}$ and $\mu_{1}$ abruptly appear at critical values of $\tchi=\sqrt{2}$ and $\tchi=1/\sqrt{3}$, respectively; for larger values of $\tchi$ the self-consistent Hartree cosine potential is zero (i.e.\ $q=0$) and the solutions $\psi_0$ and $\psi_1$ are ordinary sine waves so we have not plotted that portion of their eigenvalues. Like in \cref{fig:q-of-chi}, $\aleph=1$ and $\aleph=2$ label the two branches of the higher solutions.}
  \label{fig:mu-of-chi}
\end{figure}

\begin{figure}[t]
\includegraphics[width=\linewidth]{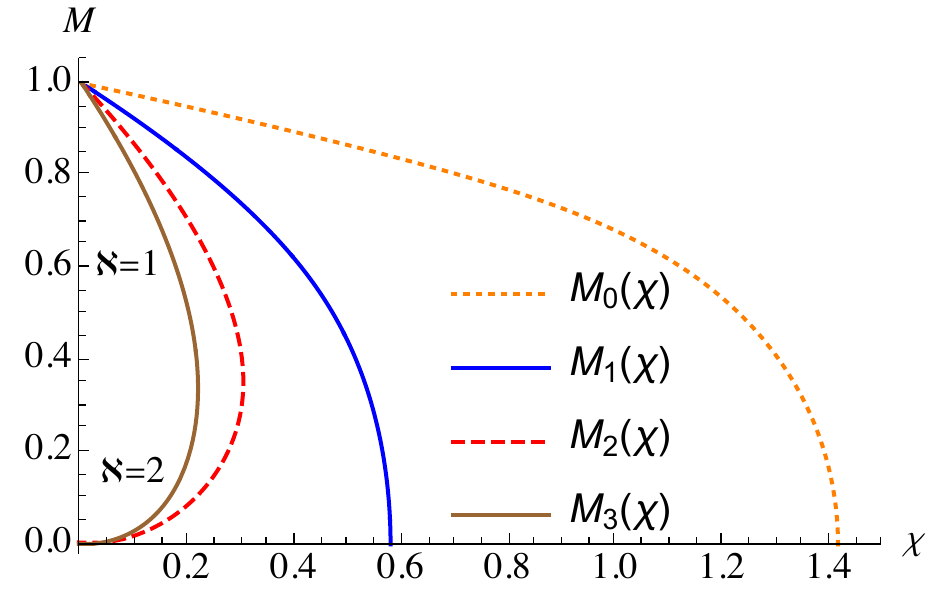}
\caption{The magnetization $M$ as a function of $\tchi$ for the first four stationary solutions. With the exception of the two lowest states, the solutions have two branches labeled by $\aleph=1$ and $\aleph=2$ like in the earlier figures. Treating the magnetization as an order parameter signifying a clustering/ordering phase transition which breaks the angular symmetry of the system, we see that this transition is second order (continuous), at least in the lowest two states. The fact that a phase transition occurs in a 1D system is a special feature of LRI.}
\label{fig:mag-of-chi}
\end{figure}

We may quantify the regime in which soliton-like solutions appear by estimating a critical value $q_n^{(c)}$ above which the stationary state is sufficiently narrow to be considered localized relative to the spatial extent of the unit circle. Demanding that $\theta_\text{turn}^+ < \pi/4$ implies that for $q$ satisfying 
\begin{equation}
  q\gtrsim q_c(n) = \frac{64}{\pi^4}(2n+1)^2\approx \frac{2}{3}(2n+1)^2,
  \label{eq:q-crit}
\end{equation}
or equivalently\footnote{Taking $M\sim \order{1}$ as $q\rightarrow \infty$. } for $\tchi$ satisfying
\begin{equation}
  \tchi \lesssim \frac{\pi^2/8}{2n+1}\approx \frac{1.23}{2n+1}  ,
  \label{eq:chi-crit}
\end{equation}
the stationary states discussed above are well localized.

From \cref{eq:turn-points} we see that for fixed $n$ the stationary-states' widths tends to zero and are localized within  an interval of size $\Delta \theta \sim \order{1/q^{1/4}}$ centered about the minimum of the meanfield potential. This suggests approximating the cosine potential as a quadratic potential well, and a thorough  analysis reveals that this can be done provided the turning point structure of the problem is preserved. The stationary solutions in this regime are  parabolic cylinder functions  \cite{Sips1949,Sips1959,Dingle1962,Frenkel2001}
\begin{equation}
  \psi_n(\theta;q)\sim \qty[\frac{\sqrt{q}}{2\pi(n!)^2}]^{1/4} 
  \begin{cases}
    D_n(\zeta) & \text{$n$  even}\\
    \cos(\tfrac12\theta)  D_n(\zeta) & \text{$n$ odd}
    \end{cases}
\end{equation}
where $\zeta=2q^{1/4}\sin(\theta/2)$  [see \cref{eq:sips-even,eq:sips-odd} for a more extensive discussion].

\subsection{Magnetization and Hartree potential depth in the strong coupling regime}
We seek to compute the magnetization 
\begin{equation}
 M(q_n)= \int_{-\pi}^{\pi} \dd \theta~ \abs{\psi_n (\theta ; q)}^2 \cos \theta
 \label{eq:mag-def}
\end{equation}
in the strong coupling regime. Expanding the wavefunction in terms of  parabolic cylinder functions, and changing variables to $\zeta=2q^{1/4}\sin(\theta/2)$, we obtain 
\begin{equation}\label{eq:mOfQ}
 M(q)= \int_{-2q^{1/4}}^{2q^{1/4}} \frac{\dd \zeta}{q^{1/4}} \qty[\frac{1-\frac{\zeta^2}{2\sqrt{q}}}{\sqrt{1-\frac{\zeta^2}{4\sqrt{q}}}}] \abs{\psi_n (\zeta ; q)}^2 .
\end{equation}
A Taylor expansion of $\qty(1-\tfrac{\zeta^2}{2\sqrt{q}})/\sqrt{1-\frac{\zeta^2}{4\sqrt{q}}}$ and knowledge of integrals of the form 
\begin{equation}
I_{n,m}^{(k)} = \int_{-\infty}^\infty \dd \zeta ~D_n(\zeta) D_m(\zeta) \zeta^k
\end{equation} 
allows one to compute asymptotic expressions for the magnetization as a function of $q$, the details of which can be found in \cref{appsec:asymptotics}. We find
\begin{equation}\label{large-q-M}
M (q_n) \sim  1-\frac{1}{\sqrt{q}}\qty[\frac{2n+1}{2}]+\order{\frac{1}{q^{3/2}}}~.
\end{equation}
This equation applies to the branches of solutions labelled by $\aleph=1$ in \cref{fig:q-of-chi}.

With this result we may now compute $\tchi_n(q)= \sqrt{4 M/q}$, which can subsequently be inverted yielding 
\begin{equation}\label{q-of-chi}
    q_n(\tchi;\aleph=1)\sim \frac{4}{\tchi^2}\qty[1-\frac{2n+1}{4}\tchi- \frac{\qty(2n+1)^2}{32}\tchi^2 ] 
\end{equation}
where, again, $\aleph=1$ denotes the branch of $q_n(\tchi)$ to which our asymptotic analysis applies. 

The above asymptotic analysis is only sensible provided that $\tchi$ is sufficiently small such that dispersive effects are weak relative to the mean field Hartree potential. Quantitatively we require that \cref{eq:chi-crit} is satisfied; for $n=0$ and $n=1$ this occurs for values of $\tchi\lesssim 1$.

\subsection{Energy in the strong coupling regime}
An analytic expression for the energy of a solitary wave can also be obtained in the small-$\chi$ limit.  The energy functional (i.e.\ Hamiltonian) for the HMF model is
\begin{equation}\begin{split}\label{energy}
    \mathcal{E}&=\int  \psi^*\qty[-\tfrac12\chi^2\partial_\theta^2]\psi \dd \theta - \tfrac12 M[\psi]^2\\
    &= \mu_n + \frac12 M[\psi_n]^2
    \end{split}
\end{equation}
where in the second equality we have used \cref{eq:NL-Mathieu}. The small-$\chi$ behavior of the magnetization $M$ is given by \cref{large-q-M,q-of-chi}
\begin{equation}
  M\sim  1 - \frac{2n+1}{4}\chi~,
\end{equation}
while the chemical potential's behaviour can be found by using the large-$q$ asymptotics of $a_n\sim -2q + (4n+2)\sqrt{q}$, which, when written in  terms of  $\chi$ gives
\begin{equation}\begin{split}
    \mu_n &=\frac{\chi^2}{8} \times a_n \sim  \frac{\chi^2}{2} \qty[-2q(\chi)+ (4n+2)\sqrt{q(\chi)}]\\
    &\sim -1 + \tfrac34(2n+1)\chi~.
\end{split}\end{equation}
Thus we have
\begin{equation}
  \mathcal{E}_n \sim -\frac12 + \frac{2n+1}{4}\chi \ ,
\end{equation}
such that the energy increases linearly with $n$ at leading order in $\chi$; a non-perturbative graph of the energy's $n$-dependence can be obtained by combining the curves plotted in \cref{fig:mu-of-chi,fig:mag-of-chi}.

\section{Linear stability analysis \label{sec:Stability}}
In order to understand whether the solitary waves analyzed in the previous two sections are stable, we study their linearized equations of motion in the soliton's rest frame. Since we identify these solutions as being solitary waves in the limit that $\chi\rightarrow 0$, and since for $n\geq2$ the solutions only exist for small values of $\chi$, we restrict our analysis to the $\aleph=1$ branch of solutions in the limit that $\chi\rightarrow0$ such that a large-$q$ expansion is justified.

We consider stability of the soliton solutions with respect to perturbations of the Bogoliubov form 
\begin{align}\label{bogo-1}
  \Psi(\theta,\tau)&=\qty[\psi_n(\theta) +  \delta\psi_n(\theta,\tau)]\e^{-\iu\mu\tau/\chi}~\\
  \delta\psi_n(\theta,\tau) &=  \sum_\alpha U_\alpha(\theta)\e^{-\iu \omega_\alpha \tau/\chi} + V_\alpha^*(\theta)\e^{\iu \omega_\alpha^* \tau/\chi}\label{bogo-2}
\end{align}
which have real frequencies if the unperturbed solution is stable, and complex frequencies if it is dynamically unstable \cite{Pethick2002}.
Substituting this ansatz into \cref{eq:GGPE}, working to first order in $U_\alpha$ and $V_\alpha^*$, and collecting terms varying in time as $\e^{-\iu \omega_\alpha \tau/\chi}$ and $\e^{\iu \omega^*_\alpha \tau/\chi}$ respectively,  one finds the following coupled equations for the normal modes
\begin{align}
  \label{eq:BdG1}
  \omega U(\theta) &= \qty(\widehat{H}_n-\mu_n)U(\theta) - \Phi(\theta) \psi_n(\theta) \\
  -\omega V(\theta) &= \qty(\widehat{H}_n-\mu_n)V(\theta) -\Phi(\theta) \psi_n(\theta)\label{eq:BdG2}
\end{align}
where we have used the reality of the  soliton, $\psi_n^*=\psi_n$ (since we are in the rest frame of the solitary wave). The subscript $\alpha$ has been left implicit,  $\widehat{H}_n=-\tfrac{1}{2}\tchi^2\partial_\theta^2 - M[\psi_n]\cos\theta$, and we define the linearized mean field potential as
\begin{equation}
  \Phi(\theta) = \int_{-\pi}^\pi \qty[U(\theta')+V(\theta')]\psi_n(\theta') \cos(\theta-\theta') \dd \theta'.
\end{equation}
It is convenient to decompose both $U(\theta)$ and $V(\theta)$ in terms of the set of Mathieu functions orthogonal to $\psi_n$ which we will denote by $\{\phi_m\}$. Note that these solutions are complete and satisfy the \emph{linear} equation
\begin{equation}
  \widehat{H}_n\phi_m = -\tfrac12\chi^2 \phi_m'' + M[\psi_n]\cos(\theta) \ \phi_m= \lambda_m\phi_m~,
\end{equation}
where $\lambda_m$ is related to the eigenvalue $a_m$ of \cref{eq:Mathieu} via $\lambda_m=\tchi^2 a_m/8$ [cf.\ \cref{eq:Self-Consistent}]. Importantly, $M[\psi_n]$ is the magnetization induced by the soliton solution $\psi_n$, and is independent of $\phi_m$.  Explicitly, we have $U (\theta)=\sum_m u_m \phi_m(\theta)$ and $V(\theta)=\sum_m v_m \phi_m(\theta)$ where $u_m$ and $v_m$ are c-numbers and $m\neq n$. This final condition ensures that our linear operator is diagonalizable \cite{Castin2001}. 

To find the frequencies  $\{\omega_\alpha\}$, we take \cref{eq:BdG1,eq:BdG2} and operate on both sides with $\int \dd\theta \phi_m(\theta)$. By virtue of the orthogonality relations  $\int \phi_m \phi_\ell \dd \theta=\delta_{m\ell}$ and $\int \phi_m\psi_n \dd \theta =0$,  this  projects out the $m^\text{th}$ element of $\{\phi_m\}$ and allows us to express \cref{eq:BdG1} in terms of the coefficients $u_\ell$ and $v_\ell$ via 
\begin{subequations}
\begin{equation}
  \omega u_m= \qty(\lambda_m-\mu_n) u_m - \sum_{\ell\neq n} F_{m\ell}\qty(u_\ell + v_\ell)
\end{equation}
\begin{equation}
  -\omega v_m= \qty(\lambda_m-\mu_n) v_m - \sum_{\ell\neq n} F_{m\ell}\qty(u_\ell + v_\ell)~,
\end{equation}
\end{subequations}
where the quantity $F_{m\ell}$ is defined as 
\begin{equation} \label{f-matrix}
 F_{m\ell}=  \mathcal{I}_{\ell,n}^C\mathcal{I}_{m,n}^C + \mathcal{I}_{\ell,n}^S\mathcal{I}_{m,n}^S~,
\end{equation}
and where the integrals $\mathcal{I}_{\ell,n}^C$ and $\mathcal{I}_{\ell,n}^S$ are defined in terms of the meanfield solutions via
\begin{subequations}
  \begin{equation}
    \mathcal{I}^C_{m,n}=\int_{-\pi}^{\pi} \phi_m(\theta)\psi_n(\theta) \cos\theta \dd \theta
  \end{equation}
  \begin{equation}
    \mathcal{I}^S_{m,n}=\int_{-\pi}^{\pi} \phi_m(\theta)\psi_n(\theta) \sin\theta \dd \theta \ .
  \end{equation}
\end{subequations}
We may interpret $F_{m\ell}$ as a matrix operator acting on the vector $(u+v)_{\ell}$ such that the equations can be cast in the form
\begin{equation}
  \label{eq:eigs-2by2}
  \omega_n \begin{pmatrix}
    u\\
    v
  \end{pmatrix} 
  =  \frac{1}{\sqrt{q}}\begin{pmatrix}
    \mathcal{D}_n - \mathcal{F}_n & -\mathcal{F}_n \\
    \mathcal{F}_n & -(\mathcal{D}_n - \mathcal{F}_n) 
  \end{pmatrix}
  \begin{pmatrix}
    u \\
    v
  \end{pmatrix},
\end{equation}
Here, $\mathcal{D}_n$ is a diagonal matrix with $\mathcal{D}_n^{m,m}=\sqrt{q}(\lambda_m-\mu_n)$ along the diagonal, while the matrix elements of $\mathcal{F}$ are related to those of $F$ via $\mathcal{F}_{m\ell}=\sqrt{q}F_{m \ell}$. Physically,  $\mathcal{D}_n$ tells us whether an orthogonal Mathieu function $\phi_m$ has a larger or smaller eigenvalue (chemical potential) as compared to the  soliton $\psi_n$. The matrix $\mathcal{F}_n$ is the mode-to-mode coupling induced indirectly by the soliton.

Explicit expressions for the matrix elements $F_{m\ell}$ and $\mathcal{D}_n^{m,m}$ can be obtained in the large-$q$ (i.e.\ small-$\chi$) regime. As outlined in \cref{appsec:someintegrals}, by making large-$q$ expansions of the integrals $\mathcal{I}^{C}_{m,n}$ and $\mathcal{I}^S_{m,n}$, then  at $\order{1/\sqrt{q}}$ only $\mathcal{I}^S_{m,n}$ contributes to $F_{m\ell}$  with the explicit (leading order) formula being 
\begin{equation}
\mathcal{I}^{S}_{m,n}\sim  \frac{1}{q^{1/4}} \qty(\sqrt{n+1}\delta_{m,n+1}+\sqrt{n}\delta_{m,n-1}) \ .
\end{equation}
Higher order terms connect states with $m=n\pm 2, m=n\pm 3$, etc. Likewise, to find the large-$q$ behavior of $\mathcal{D}_n$, we can use the large-$q$ asymptotic formula for the eigenvalues of the Mathieu equation \cite{Olver2010}
\begin{align}
  \label{eigs-large-q1}
  \mu_n &\sim \frac{\chi^2(q_n)}{8}[-2 q + 2(2n+1)\sqrt{q} + \frac18((2n+1)^2+1)]\\
  \lambda_m &\sim  \frac{\chi^2(q_n)}{8}[-2 q + 2(2m+1)\sqrt{q} + \frac18((2m+1)^2+1)] . \label{eigs-large-q2}
\end{align}
Notice that both $\lambda_m$ and $\mu_n$ are multiplied by the same value of $\chi^2$ which corresponds to the soliton's self-consistent depth parameter $q_n$. At leading order this implies that $\mathcal{D}_n^{m,\ell}\sim 2(m-n)\delta_{\ell,m}$, and that all of $\mathcal{F}_n$'s non-vanishing entries are contained within a $2\times2$ block composed of $\ell,m=n\pm1$ (unless n=0) given by 
\begin{equation}
  \begin{pmatrix}
    n  & \sqrt{n(n+1)}\\
    \sqrt{n(n+1)} & n+1 
  \end{pmatrix}~.
\end{equation}
This implies that in the strong-coupling regime the perturbations about the soliton are \emph{weakly interacting}, with the exception of the two Mathieu-modes whose eigenvalues are closest to the soliton's chemical potential. These two modes actually  couple with a strength that is $\order{n}$ such that solitons with a  higher number of nodes mediate stronger interactions than those with fewer nodes.  By contrast, if $n=0$  (i.e.\ if we are perturbing around the lowest energy soliton), then inter-mode coupling does not exist at  $\mathcal{O}(1/\sqrt{q})$ and  $\mathcal{F}_n$ is diagonal at leading order, having all vanishing entries except for $\mathcal{F}_n^{11}=1$. We focus now on $n\geq1$ after which we will return to $n=0$ as a special case.

By re-writing \cref{eq:eigs-2by2} in terms of $u+v$ and $u-v$, it can be easily seen that the eigenvalues of the above matrix equation are determined by the condition that 
\begin{equation}
  \textrm{Det}\qty[q\omega_n^2\mathds{1} - (\mathcal{D}_n^2 - 2\mathcal{F}_n\mathcal{D}_n)]=0.
\end{equation}
The spectrum of $\mathcal{D}_n^2-2\mathcal{F}_n\mathcal{D}_n$ is the same as that of $\mathcal{D}_n^2$ except for the two eigenvalues that are determined (for $n\geq1$)  by the $2\times2$ matrix
\begin{equation}\label{2times2}
  \begin{pmatrix}
    4 n+4 & -4 \sqrt{n (n+1)} \\
    4 \sqrt{n (n+1)} & -4 n
  \end{pmatrix},
\end{equation}
whose eigenvalues are independent of $n$ and given by $\lambda_1=0$ and $\lambda_2=4$. We thereby find that the entire spectrum is positive, indicating stability, except for one entry which is marginal at this order, being neither positive or negative. 

To elucidate the stability of the mode corresponding to $\lambda_{1}$ we must calculate sub-dominant corrections to  $\mathcal{I}^S$, and $\lambda_m-\mu_n$. This can be achieved using first-order perturbation theory for which we need the eigenvector corresponding  to $\lambda_1$ at leading order, which is given by
\begin{equation}
  v_1=(\sqrt{n/(2n+1)},  \sqrt{(n+1)/(2n+1)})^T~. 
\end{equation}
If we denote the sub-dominant corrections as  $\delta \mathcal{D}_n$ and $\delta \mathcal{F}_n$ (such that $\mathcal{D}_n\rightarrow \mathcal{D}_n + q^{-1/2}\delta \mathcal{D}_n$ and $\mathcal{F}_n\rightarrow \mathcal{F}_n+ q^{-1/2}\delta\mathcal{F}_n$) then, the first-order perturbative correction to the eigenvalue is given by
\begin{equation}
  \lambda_1\sim \frac{1}{\sqrt{q}}v_1^T \qty[\{\mathcal{D}_n,  \delta\mathcal{D}_n\} - 2(\delta\mathcal{F}_n \mathcal{D}_n +  \mathcal{F}_n\delta \mathcal{D}_n)] v_1~,
\end{equation}
where  the curly-braces denote an anti-commutator. Using \cref{eigs-large-q1,eigs-large-q2} we find for $\delta\mathcal{D}_n$
\begin{equation}
  \delta{D}_n=\begin{pmatrix}
  \frac{1}{4} (10 n+4)  & 0\\
  0 & \frac14 (-6 - 10 n) ~.
  \end{pmatrix}
\end{equation}
The relevant matrix elements of $\mathcal{F}$ are  $\mathcal{F}_{\ell m}$ with $\ell,m=n\pm 1$ which can be represented as a $2\times2$ matrix
\begin{equation}\label{delta-f}
  \delta\mathcal{F}=
  \begin{pmatrix}
    -\frac34 n^2 & -\frac{3}{8}\sqrt{n(n+1)} (2 n+1) \\
    -\frac{3}{8} \sqrt{n(n+1)} (2 n+1) & -\frac34(1+n)^2
  \end{pmatrix}
\end{equation}
such that
\begin{equation}
  \lambda_1\sim  \frac{1}{\sqrt{q}}\times \frac{n(1+n)}{2n+1}\qq{for} n\geq1
\end{equation}
Thus, for $n\geq1$ all of the eigenvalues $\lambda$ are positive definite, such that  $\omega_m$ is always real, and the solitons are dynamically stable. 
The typical frequencies are given by $\omega_m\sim \frac{2}{\sqrt{q}}|m-n|\sim \chi |m-n|$, while the smallest frequency is parametrically smaller being given by $\omega_1=\pm \sqrt{\lambda_1/q}\sim \mathcal{O}({\chi^{3/2}})$.

As mentioned above, the case of the $n=0$ is different. Here, the zeroth  eigenvector  is  $v_1=(1,0)^T$ and it turns out that its eigenvalue is small  $\lambda_1\sim \order{1/q}$, If we define the matrix $M=\mathcal{D}_n^2 -2\mathcal{F}_n\mathcal{D}_n$, then we have that
\begin{equation}
  M= \begin{pmatrix}
    M_{11} & 0 & M_{13} \\
    0 & M_{22} & 0 \\
    M_{31}& 0 & M_{33}
  \end{pmatrix}~.
\end{equation}
where the zeros stem from the fact that $\mathcal{I}^C_{0,m}=0$ if $m$ is odd. As we will soon see, $M_{13}$ and $M_{31}$ are both $\order{1/\sqrt{q}}$ while $M_{33}$ is $\order{1}$, and we can therefore calculate the correction to $\lambda_1$ within  second order perturbation theory
\begin{equation}\label{2ndOrderPert}
  \lambda_1 \sim M_{11} + \frac{M_{13} M_{31}}{M_{11}-M_{33}} + \order{q^{-3/2}}  \qq{for} n=0~.
\end{equation}
Explicit formula for $M_{11}$, $M_{13}$, and $M_{31}$ in terms of $\mathcal{I}^S_{0,m}$ and $\Delta_{m}=\lambda_m-\mu_0$ are given by
\begin{align}
  M_{11} &= q\Delta_1^2 -2q\Delta_1\mathcal{I}_{0,1}^S\mathcal{I}_{0,1}^S
  \label{Meq1}\\
  M_{13} &= -2q\Delta_3 \mathcal{I}_{0,1}^S\mathcal{I}_{0,3}^S \label{Meq2}\\
  M_{31} &= -2q\Delta_1\mathcal{I}_{0,1}^S\mathcal{I}_{0,3}^S \label{Meq3}\\
  M_{33} &= q\Delta_3^2 -2q\Delta_3\mathcal{I}_{0,3}^S\mathcal{I}_{0,3}^S\label{Meq4}
\end{align}
The large $q$ behavior of $\Delta_m$ can be found by using Eq.\ 28.8.1  of  \cite{Olver2010} and the next-to-next-to leading  order expression for $\chi(q_n)$ given in \cref{chi-NNLO}. To compute $\mathcal{I}^S_{0,m}$ we make use of Eqs.\ (28.8.3) to (28.8.7) from \cite{Olver2010} at next-to-next-to leading order accuracy. At the  level of  accuracy required  to compute $\lambda_1$ the results are
\begin{align}
  q^{1/4}\mathcal{I}^S_{0,1}&\sim 1-\frac{3}{8\sqrt{q}} - \frac{19}{256 q}\\
  q^{1/4}\mathcal{I}^S_{0,3}&\sim -\frac{3\sqrt{3/2}}{8\sqrt{q}} \\
  \sqrt{q}\Delta_1&\sim 2- \frac{3}{2\sqrt{q}} + \frac{3}{16 q} \\
  \sqrt{q}\Delta_3&\sim 6 ~,
\end{align}
Substituting these expressions into \cref{Meq1,Meq2,Meq3,Meq4} we find
\begin{align}
  M_{11}&\sim \frac{13}{32 q} + \order{\frac{1}{q^{3/2}} } \\
  M_{13}&\sim \frac{9 \sqrt{3/2}}{2\sqrt{q}} + \order{\frac1q}\\
  M_{31}&\sim \frac{3 \sqrt{3/2}}{2 \sqrt{q}} + \order{\frac1q}\\
  M_{33}&\sim 36 + \order{\frac1{q^{1/2}}}~.
\end{align}
Then, using \cref{2ndOrderPert} we arrive at
\begin{equation}
  \lambda_1 \sim \frac{1}{8 q} \qq{for} n=0~.
\end{equation}
Thus,  the  $n=0$  soliton  is also stable  (as it must be since it is the lowest energy stationary state). Curiously, whilst all of the other  solitons' smallest normal mode frequencies  are $\mathcal{O}(\chi^{3/2})$,  the  ground state's lowest lying normal mode frequency is  actually $\mathcal{O}(\chi^2)$  since  $\omega_1 =\sqrt{\lambda_1/q}\sim \mathcal{O}(\chi^2)$.
\section{Conclusions \label{sec:Conclusions}}

We have shown that the GGPE for the HMF model admits exact solutions in the form of Mathieu functions complemented by a self-consistency condition on the depth of the Hartree potential they generate. These solutions can be boosted to finite speeds (providing the phase associated with the flow satisfies the periodic boundary conditions). In the strong coupling regime ($\tchi\ll 1$) the solutions can be arbitrarily highly localized and thus we interpret them as bright solitons. A linear stability analysis in the same strong coupling regime shows that they are stable against perturbations when the interactions are attractive. 
 
The fact that the solutions: 1) arise in a periodic potential, 2) are localized, and 3) come in towers with increasing numbers of nodes, means that they have properties in common with gap solitons \cite{Louis03}. However, the periodic potential in the HMF case is self-generated, whereas in the standard gap soliton case the periodic potential is imposed externally. Furthermore, in contrast to the  bright soliton solutions of the standard GPE which are stabilized by a $\abs{\psi}^2\psi$ nonlinearity, the HMF model's solitons are stabilized by a nonlocal nonlinearity $\Phi[\psi](\theta)=M[\psi]\cos\theta$, where the depth is given by the magnetization $M[\psi]$, see \cref{eq:GGPE-b}.

The approach followed in this work not only allows us to identify solitary wave solutions, but also to find and classify all possible stationary states for both attractive and repulsive interactions at arbitrary coupling strength (we focused on the attractive case). Furthermore, in the strong coupling limit the self-consistency condition can be developed analytically in an asymptotic series so as to provide a completely explicit analytic solution.
Given that exact solutions of nonlinear models are few and far between, this illustrates once again that the HMF model is rather special even though it is not thought to be integrable.

One possible reason for the existence of a richer family of solutions (i.e.\ the tower of solutions) in comparison to the standard short-range attractive system, where the GPE supports a single nodeless bright soliton [i.e.\ the fundamental soliton given in Eq.\ (\ref{eq:fundamental})], is that the self-consistency condition in the latter case is much more restrictive: both the depth and shape must match.  By contrast, in the HMF case the meanfield potential is generated by the coherent addition of the microscopic cosine (XY) interactions and thus inherits a cosine form where only the depth needs to be self-consistently determined. We conjecture that this coherent addition across the sample is a generic feature of LRI, and suggests that such systems deserve closer examination as a potential setting for solitons.

 An interesting future direction of research would be to study collisional properties of the bright solitons (solitary waves) identified in this work and to determine if they are true solitons (i.e.\ do they collide elastically). As sketched in the introduction, systems with LRI can be expected to behave similarly to integrable systems. A numerical study of soliton dynamics in the HMF model would be a natural testing ground for this idea. 

Another  extension of the present work concerns the quantum phase transition predicted by the HMF model's GGPE due to a spontaneous breaking of translational invariance at the critical value of $\tchi=\sqrt{2}$ \cite{Chavanis2011}. The GGPE does not include the effects of quantum fluctuations, and these may inhibit this spontaneous symmetry breaking.  However, our exact solutions can serve as the building blocks of more sophisticated quantum states that are required for studying the role of quantum fluctuations. These effects will be discussed elsewhere \cite{Ryan-inprep}.

\section{Acknowledgements}
We thank Dmitry Pelinovsky, Robert Dingwall, Wyatt Kirkby, and Thomas Bland for helpful discussions. This research was funded by the Natural Sciences and Engineering Research Council of Canada (NSERC) and the Government of Ontario. Support is also acknowledged from the Perimeter Institute for Theoretical Physics. Research at the Perimeter Institute is supported by the Government of Canada through the Department of Innovation, Science and Economic Development and by the Province of Ontario through the Ministry of Research and Innovation.

\appendix
\crefalias{section}{appsec}

\section{Asymptotic analysis of the self-consistent magnetization}
\label{appsec:asymptotics}

We are  interested in calculating the magnetization 
\begin{equation}\begin{split}
\label{eq:mag-int}
M_n(\tchi)&=\int_{-\pi}^{\pi} \dd \theta \abs{\psi_n(\theta ; \tchi)}^2 \cos\theta\\
\nonumber &= \int_{-\pi}^{\pi} \dd \theta \abs{\psi_n(\theta ; \tchi)}^2 \qty(1-2\sin^2\tfrac\theta2)\\
          &= 1- 2\int_{-\pi}^{\pi} \dd \theta \abs{\psi_n(\theta ; \tchi)}^2 \sin^2\tfrac\theta2
\end{split}\end{equation}
in the strong coupling regime where $\tchi\ll 1$. We must consider the cases of $n$ even and $n$ odd separately due to their definition in terms of either even or odd Mathieu functions which we repeat here for convenience
\begin{equation}
\psi_n(\theta ; \tchi,\aleph)=\frac{1}{\sqrt{\pi}}\begin{cases}
\hspace*{0.9em}\textrm{ce}_{n}\qty[\frac{\theta-\pi}{2} ;q_n(\tchi ;\aleph)]\hspace*{-0.2em} & n~\text{even} \\
\textrm{se}_{n+1}\qty[\frac{\theta-\pi}{2}~;q_n(\tchi ; \aleph)]\hspace*{-0.2em}
 & n~\text{odd} \, .
\label{eq:psi-n-def-appendix}
\end{cases}
\end{equation}
The strong coupling regime is equivalent to $q\gg 1$ in the conventional Mathieu equation. When applied to the HMF model, this means that one can study the regime in which $q\gg 1$, compute the magnetization and by extension $\tchi(q)=\sqrt{4 M(q)/q}$, and then invert this expression to find how $q$ depends on $\tchi$. 

Consequently, it is convenient for computational purposes to consider $q$, rather than $\tchi$ as a fixed parameter, and later invert the relationship between them as described above. Integrals such as \cref{eq:mag-int} are conveniently analyzed by transforming to the coordinate $\zeta=2q^{1/4}\sin\tfrac{\theta}{2}$; doing so we find [as in \cref{eq:mOfQ}]
\begin{equation}
M_n(q)= 1 - \frac{1}{2\sqrt{q}}\int_{-2q^{1/4}}^{2q^{1/4}} \frac{\dd \zeta}{q^{1/4}} \abs{\psi_n (\zeta ; q)}^2 \frac{\zeta^2}{\sqrt{1 -\frac{\zeta^{2}}{4\sqrt{q}}} },
\label{eq:attack-int}
\end{equation}

For the purposes of obtaining an asymptotic series in $1/\sqrt{q}$ we may extend the limits of integration to $\pm \infty$,  
\begin{equation}
M_n(q)\sim 1 - \frac{1}{2\sqrt{q}}\int_{-\infty}^{\infty} \frac{\dd \zeta}{q^{1/4}} \abs{\psi_n (\zeta ; q)}^2 \frac{\zeta^2}{\sqrt{1 -\frac{\zeta^{2}}{4\sqrt{q}}} },
\label{eq:attack-int}
\end{equation}
where $\sim$ denotes an asymptotically small error as $q\rightarrow\infty$ \cite{Olver2010}. We may then expand the square root in the denominator in a Taylor series, and use the large-$q$ behavior of the stationary solutions $\psi_n(\zeta ; q)$.

\begin{figure}[t]
  \includegraphics[width=\linewidth]{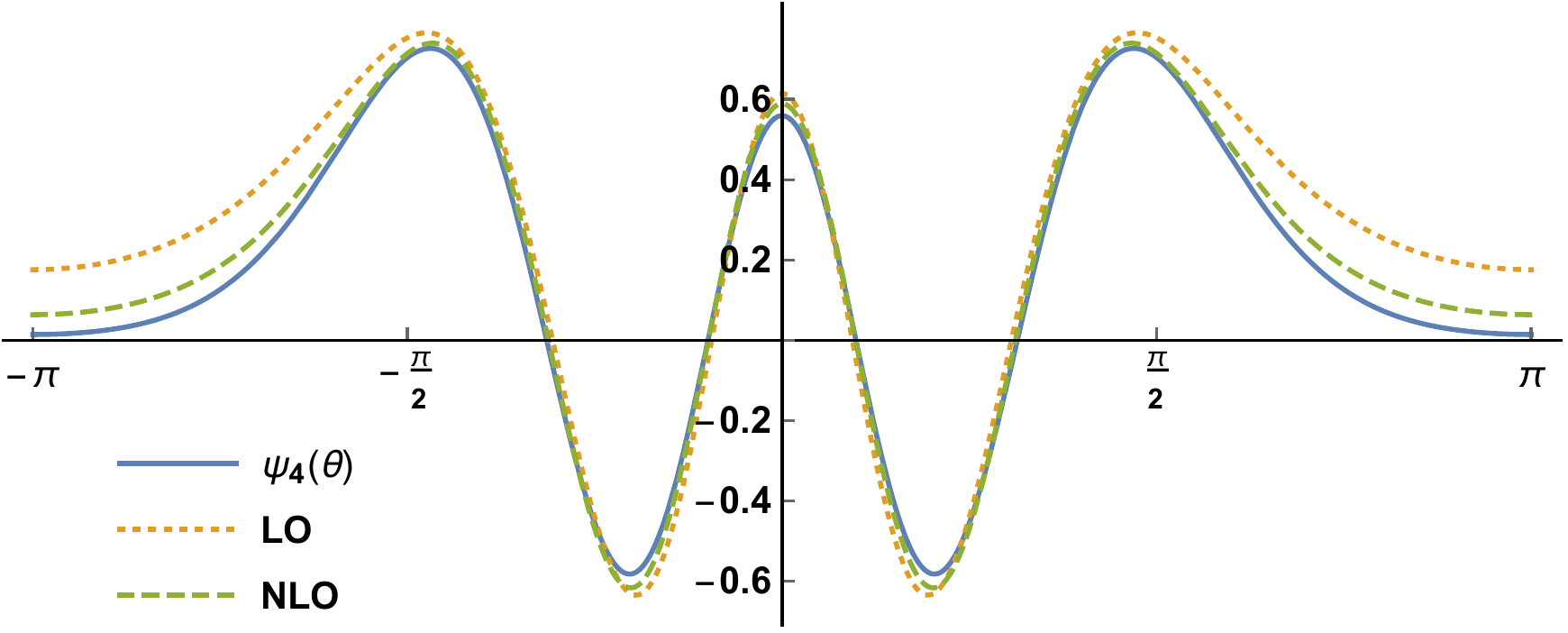}\\
  \caption{Comparison of the $n=4$ exact solitary wave $\psi_n(\theta,q)$ for $q=\tfrac23 (2n+1)^2=54$. LO denotes the leading-order Sips expansion \cref{eq:sips-even,eq:sips-odd}, while NLO denotes the approximation including next-to-leading order terms. As we can see, the Sips expansion becomes accurate even at only moderately large values of $q$. For $q=(2n+1)^2=81$ the exact and NLO curves become virtually indistinguishable.}
  \label{sips-pic}
\end{figure}

In the limit of large $q$ both $\ce_n(z ; q)$ and $\se_n(z ; q)$ can be expanded in terms of Parabolic cylinder functions \cite{Sips1949,Sips1959,Dingle1962,Frenkel2001,Olver2010}. This is easy to understand, since for $n$ fixed and $q\rightarrow \infty$ the classical turning points coalesce at the minimum of the cosine potential. Therefore the states are constrained to live arbitrarily close to the potential's minimum and a harmonic approximation is justified. To ensure that the turning point structure is maintained the expansion is carried out using $\xi=2q^{1/4}\cos z$ rather than  the naive choice of $z-\pi/2$ . Explicitly, Sips's expansion of the Mathieu functions in terms of parabolic cylinder functions assumes the form 
\cite{Olver2010} 
\begin{align}
\label{eq:sips-even}
\ce_n(z ; q)\sim& \widehat{C}_n(q) \qty[ \widehat{U}_n(\xi ; q) + \widehat{V}_n(\xi ; q)] \\
 \se_{n-1}(z ; q)\sim& \widehat{S}_n(q) \sin(z) \qty[\widehat{U}_n(\xi ; q)-\widehat{V}_n(\xi ; q)]
 \label{eq:sips-odd}
\end{align}
where $U_m(\xi ; q)$ and $V_m(\xi ; q)$ are given at next-to-leading order [i.e.\ suppressing terms of $\mathcal{O}(q)$] by
\begin{align}
  \label{eq:sips-u}
  \begin{split}
    \widehat{U}_n(\xi ; q)&\sim D_n(\xi)\\
    & ~~+ \frac{1}{64\sqrt{q}} \qty[\frac{n!}{(n-4)!}D_{n-4}(\xi) -D_{n+4}(\xi)]
  \end{split}\\
    \label{eq:sips-v}  \widehat{V}_n(\xi ; q)&\sim \frac{1}{16\sqrt{q}} \bigg[n(1-n)D_{n-2}(\xi) -  D_{n+2}(\xi)   \bigg] 
\end{align}
where $D_n=D_n(\xi)$ are the parabolic cylinder functions of order $n$, and the normalization constants are given by
\begin{align}
\label{eq:consts1}
 \widehat{C}_n(q)\sim& \qty[\frac{\pi \sqrt{q}}{2(n!)^2}]^{1/4}\qty[ 1 + \frac{(2n+1)}{8\sqrt{q}} +  \order{\frac{1}{q}} ]^{-1/2}\\
 \widehat{S}_n(q)\sim& \qty[\frac{\pi \sqrt{q}}{2(n!)^2}]^{1/4}\qty[1 -\frac{(2n+1)}{8\sqrt{q}} + \order{\frac{1}{q}}]^{-1/2}\hspace{-5pt}.
 \label{eq:consts2}
\end{align}
The accuracy of the Sips expansion even for relatively modest values of $q$ is illustrated in \cref{sips-pic}.

Since we have a series solution to $\psi_n(\zeta ; q)$ in terms of parabolic cylinder functions, and we are interested in computing integrals  $\int \abs{\psi_n}^2 \zeta^k \dd \zeta$, we ultimately require knowledge of integrals of the form 
\begin{align}
  I_{m,n}^{(k)}=\int_{-\infty}^\infty D_m(\xi) D_n(\xi)\xi^k \dd \xi.
\end{align}
Then using the  recursion relation \cite{Frenkel2001}
\begin{equation}
  I^{(k)}_{m,n}=I^{(k-1)}_{m+1,n} + I^{(k-1)}_{m,n+1} - (k-1) I^{(k-2)}_{m,n},
\end{equation}
we arrive at the following identities
\begin{align}
  I_{m,n}^{(0)}&=n!\sqrt{2\pi}\times \delta_{m,n}  \label{eq:int1} \\
  I^{(1)}_{m,n}&=n!\sqrt{2\pi}\times \qty[(n+1)\delta_{m,n+1}+ \delta_{m+1,n}] \label{eq:int2}\\
  I_{m,n}^{(2)}&=n!\sqrt{2\pi}\times [(n+2)(n+1)\delta_{m,n+2} \label{eq:int3} \\ &\notag \quad \quad \quad\quad\quad\quad  + (2n+1)\delta_{m,n} + \delta_{m+2,n}]\\
  I_{n,n}^{(4)}& = n!\sqrt{2\pi}\times 3(2n^2+2n+1) \label{eq:int4}.
\end{align}
Armed with these details, we may now attack the integral in \cref{eq:attack-int}. First, we note that by \cref{eq:psi-n-def-appendix} that in using \cref{eq:sips-even,eq:sips-odd} we must make the substitution $z\rightarrow (\theta+\pi)/2$, which in turn implies that $\xi\rightarrow \zeta = 2 q^{1/4} \sin (\theta/2)$ and $\sin z\rightarrow \cos(\theta/2)$. Expressing all functions in terms of $\zeta$, and then performing a Taylor expansion yields 
\begin{widetext}
  \begin{equation}    
    M_n(q)\sim 1- \int_{-\infty}^{\infty}\frac{\dd\zeta}{q^{1/4}} \qty[\frac{1}{\sqrt{\pi}}]^2 \bigg[\widehat{U}_n(\zeta)\pm \widehat{V}_n(\zeta) \bigg]^2 \qty[\qty(\frac{\pi \sqrt{q} }{2(n!)^2})^{1/4} ]^2 \qty(1\mp \frac{2n+1}{8\sqrt{q}})\times \qty( \frac{1}{2\sqrt{q}} \zeta^2  \pm \frac{1}{16 q}\zeta^4),
    \label{eq:mag-expanded}
  \end{equation}
  where the upper sign corresponds to $n$ even and the lower sign to $n$ odd. Using \cref{eq:int1,eq:int2,eq:int3}, the integral may then be expressed in terms of $I_{m,n}^{(k)}$ as 
  \begin{equation}
    \begin{split}
      M_n(q)&\sim1- \frac{1}{n!\sqrt{2\pi}}\qty(\frac{1}{2\sqrt{q} } I_{n,n}^{(2)}
      \pm \frac{1}{16 q}\qty[ I_{n,n}^{(4)} +n(1-n)I_{n,n-2}^{(2)}-I_{n,n+2}^{(2)} -(2n+1)I_{n,n}^{(2)}  ])+\order{\frac{1}{q^{3/2}}}\\
      &= 1-\frac{1}{2\sqrt{q}}[2n+1] \mp \frac{1}{16 q}\qty[ 3(2n^2 +2n+1)   + n(1-n)-(n+1)(n+2)-(2n+1)^2]+\order{\frac{1}{q^{3/2}}}\\
      &= 1-\frac{1}{2\sqrt{q}}[2n+1]+\order{\frac{1}{q^{3/2}}}.
    \end{split}
  \end{equation}
\end{widetext}
Having obtained the magnetization in terms of the auxillary parameter $q$, we now compute $\tchi(q)$ and find at next-to-next-to leading order
\begin{align}\label{chi-NNLO}
  \chi(q_n)&=\sqrt{\frac{4 M_n(q)}{q}}\\
  \notag  &\sim \frac{2}{\sqrt{q}}\qty[1-\frac{1}{\sqrt{q}}\frac{2n+1}{4} - \frac{1}{q}\frac{(2n+1)^2}{32}],
\end{align}
which may in turn be inverted to find
\begin{equation}
  q_n(\tchi ; \aleph=1)\sim \frac{4}{\tchi^2}\qty[1-\frac{2n+1}{4}\tchi- \frac{\qty(2n+1)^2}{32}\tchi^2 ]
  \label{eq:q-self-cons}
\end{equation}
as claimed in the main text. The branch labeled by $\aleph=1$ is the appropriate one since we took the auxiliary variable $q$ to be large.

\section{Definition and asymptotics of the integrals $\mathcal{I}^C_{m,n}$ and $\mathcal{I}^S_{m,n}$ \label{appsec:someintegrals}}
The following integrals appear in \cref{f-matrix}
\begin{align}
  \mathcal{I}^C_{m,n}&=\int_{-\pi}^{\pi} \phi_m(\theta)\psi_n(\theta) \cos\theta \dd \theta\\
  \mathcal{I}^S_{m,n}&=\int_{-\pi}^{\pi} \phi_m(\theta)\psi_n(\theta) \sin\theta \dd \theta,
\end{align}
where $\psi_n$ is the stationary state [given by a Mathieu function with depth parameter $q(\chi)$] around which fluctuations take place, and $\phi_m$ is a Mathieu function orthogonal\footnote{The functions  $\{\psi_m\}$  satisify the (rescaled) linear Mathieu equation $-\frac{\chi^2}{2}\phi_m'' - M[\psi_n]\cos\theta\phi_m = \lambda_m \phi_m$.} to $\psi_m$.

Note that both $\psi_n$ and $\psi_m$, being Mathieu functions,  are strictly even or odd, and so $\mathcal{I}^C_{m,n}=0$ identically if $m+n$ is odd, while $\mathcal{I}^S_{m,n}=0$ identically if $m+n$ is even. We can compute these integrals order-by-order in $1/\sqrt{q}$ by re-expressing them in terms of $\zeta=2q^{1/4}\sin\tfrac\theta2$. For the sine-integral we find,  
\begin{equation}
  \begin{split}
  \mathcal{I}^S_{m,n}= \int_{-2q^{1/4}}^{2q^{1/4}} &\qty[\frac{\dd \zeta }{q^{1/4}\qty(1-\frac{\zeta^2}{4\sqrt{q}})}] \phi_m(\zeta ; q)\psi_n (\zeta ; q)\\
  & \quad\quad\quad\quad\times \qty[\frac{\zeta}{ q^{1/4}}\sqrt{1-\frac{\zeta^2}{4\sqrt{q}}}].
  \end{split}
\end{equation}
Next, using the asymptotic formulae for Mathieu functions, \cref{eq:sips-even,eq:sips-odd}, and working at leading order in $1/\sqrt{q}$ we find
\begin{equation}\begin{split}
    \mathcal{I}^S_{m,n}&\sim \int_{-2q^{1/4} }^{ 2q^{1/4} } \dd \zeta \frac{\zeta}{\sqrt{q}}  \qty[\frac{1}{\sqrt{\pi}} \qty(\frac{\pi\sqrt{q} }{2(n!)^2})^{1/4} D_n(\zeta)] \\
    &\quad\quad\quad\quad\quad\quad\times \qty[ \frac{1}{\sqrt{\pi}} \qty(\frac{\pi\sqrt{q} }{2(m!)^2})^{1/4}D_m(\zeta)] \\
     &=\frac{1}{q^{1/4}} \frac{1}{\sqrt{2\pi m! n!}} I^{(1)}_{n,m} \\
     &=\frac{1}{q^{1/4}}\qty[\sqrt{n+1}\delta_{m,n+1} + \sqrt{n}\delta_{m+1,n}],
\end{split}\end{equation}
where we have made use of \cref{eq:int2} to move between the second and third equalities.
For \cref{delta-f} we require $\mathcal{I}^S_{m,n}$ with $m=n\pm1$ at next-to-leading order, to calculate $\delta \mathcal{F}$. These are given by
\begin{align}
  \mathcal{I}^S_{n+1,n} &=q^{1/4}\qty[ \sqrt{n+1} -\frac{3(n+1)}{8\sqrt{q}}]\\
   \mathcal{I}^S_{n-1,n}&=q^{1/4}\qty[ n -\frac{3n}{8\sqrt{q}} ]~.
\end{align}
Next, turning our attention to the cosine integral we find
\begin{equation}
  \begin{split}
  \mathcal{I}^C_{m,n}= \int_{-2q^{1/4}}^{2q^{1/4}} &\qty[\frac{\dd \zeta }{q^{1/4}\qty(1-\frac{\zeta^2}{4\sqrt{q}})}] \phi_m(\zeta ; q)\psi_n (\zeta ; q)\\
  & \quad\quad\quad\quad\times \qty[1-\frac{\zeta^2}{2\sqrt{q}}].
  \end{split}
\end{equation}
When re-expressed in terms of parabolic cylinder functions, we see that the $\order{1}$ piece vanishes ($\int D_m D_n \dd \zeta =0$ for $m\neq n$), and the remaining integral ($\int D_m D_n \zeta^2/\sqrt{q} \dd \zeta$) is $\order{1/\sqrt{q}}$. This implies that
\begin{equation}
  I^C_{m,n}\sim \order{\frac{1}{\sqrt{q}}}~.
\end{equation}
This is sub-dominant to the sine integral $I^S_{m,n}\sim \mathcal{O}(1/q^{1/4})$ and so can be neglected at leading order as claimed in the main text. 

\vfill
\vfill
\pagebreak

\bibliography{soliton-ryan2.bib}

\end{document}